\documentclass[%
reprint,
superscriptaddress,
showpacs,
amsmath,amssymb,
prl,
]{revtex4-2}
\usepackage{graphicx}% Include figure files
\usepackage{dcolumn}% Align table columns on decimal point
\usepackage{bm}% bold math
\usepackage{CJKutf8}
\usepackage{color}
\usepackage{amssymb}
\usepackage{amsfonts}
\usepackage{palatino}
\usepackage[colorlinks,urlcolor=blue,linkcolor=blue,citecolor=blue,anchorcolor=blue]{hyperref}
\makeatletter

\begin{document}

\begin{CJK*}{UTF8}{gbsn}

\title{Chiral bulk solitons in photonic graphene with decorated boundaries}% Force line breaks with \\
\author{Shuang Shen (沈双)}
\affiliation{Key Laboratory for Physical Electronics and Devices, Ministry of Education, School of Electronic Science and Engineering, Xi'an Jiaotong University, Xi'an 710049, China}
\author{Ce Shang (尚策)}
\email{shangce@aircas.ac.cn}
\affiliation{Aerospace Information Research Institute, Chinese Academy of Sciences, Beijing 100094, China}
\author{Yongdong Li (李永东)}
\affiliation{Key Laboratory for Physical Electronics and Devices, Ministry of Education, School of Electronic Science and Engineering, Xi'an Jiaotong University, Xi'an 710049, China}
\author{Yiqi Zhang (张贻齐)}
\email{zhangyiqi@xjtu.edu.cn}
\affiliation{Key Laboratory for Physical Electronics and Devices, Ministry of Education, School of Electronic Science and Engineering, Xi'an Jiaotong University, Xi'an 710049, China}
\date{\today}

\begin{abstract}

We propose a chiral bulk soliton in a nonlinear photonic lattice with decorated boundaries, presenting a novel approach to manipulate photonic transport without extensive bulk modifications. Unlike traditional methods that rely on topological edge and corner modes, our strategy leverages the robust chiral propagation of bulk modes.  By introducing nonlinearity into the system, we find a stable bulk soliton,  akin to the topological valley Hall effects. The chiral bulk soliton exhibits remarkable stability; the energy does not decay even after a long-distance propagation; and the corresponding Fourier spectrum confirms the absence of inter-valley scattering indicating a valley-locking property. Our findings not only contribute to the fundamental understanding of nonlinear photonic systems but also hold significant practical implications for the design and optimization of photonic devices.

\end{abstract}
%\keywords{Suggested keywords}%Use showkeys class option if keyword
                              %display desired
\maketitle

\end{CJK*}

Topological materials are celebrated for their robust transportation of edge states, a feature that is fundamentally determined by the bulk topological properties and reflected in their topological invariants~\cite{hasan.rmp.82.3045.2010,qi.rmp.83.1057.2011}. Within this emerging field, valleytronics offers additional control of the degree of freedom of the valley~\cite{schaibley.nrm.1.16055.2016,vitale.small.14.1801483.2018}, enabling unidirectional and robust transport of the valley Hall edge state of the valley~\cite{drouot.aim.368.107142.2020,noh.prl.120.063902.2018,zhong.ap.3.056001.2021,tang.oe.29.39755.2021,ren.nano.10.3559.2021}. Traditionally, altering the transport of topological states requires significant structural changes for all unit cells within the bulk. However, a novel strategy inverts this paradigm. 
By manipulating the edge structures, one can induce chiral transport phenomena within the bulk itself~\cite{wang.nc.13.5916.2022,zhang.prl.132.086302.2024}. 
Considering a topologically trivial two-dimensional hexagonal slab confined by hard-wall boundaries, it can engender chiral bulk anomaly within a pseudogap (finite-size gap) opened as a consequence of the finite-size effect~\cite{shen.2dm.4.035014.2017}. In this approach, the translation symmetry is broken, which means an ill definition of the topological invariant. 
Yet, the valley-locking property remains an intrinsic characteristic, 
transitioning the valley Hall edge state to the \textit{valley Hall bulk state}, thereby opening new avenues in the field of valleytronics.

Similar to valley Hall edge state, the valley Hall bulk states are chiral bulk states robust against weak disorders, as the inter-valley scattering is sufficiently small due to the large separation of the valleys in the momentum space. 
However, the finite-size gap is relatively small, making the chiral bulk states susceptible to hybridization with traditional bulk states. 
In this context, we propose a chiral bulk soliton within the framework of a nonlinear photonic lattice, characterized by its distinctively decorated boundaries.  
The photonic lattice can be fabricated in fused silica by using the femtosecond laser direct writing technique~\cite{rechtsman.nature.496.196.2013,kirsch.np.17.995.2021,arkhipova.sb.68.2017.2023,ren.light.12.194.2023,li.ap.4.024002.2022,wang.light.13.130.2024},  offering a unique testbed for the realization of different lattice structures, 
such as square~\cite{arkhipova.sb.68.2017.2023}, Lieb~\cite{vicencio.prl.114.245503.2015,mukherjee.prl.114.245504.2015}, 
honeycomb~\cite{plotnik.nm.13.57.2014}, Kagome~\cite{lang.pra.107.023509.2023}, disclination~\cite{ren.light.12.194.2023}, and fractal lattices~\cite{xu.np.15.703.2021,biesenthal.science.376.1114.2022,li.light.12.262.2023}, 
and various types of nontrivival phenomena, such as Floquet~\cite{rechtsman.nature.496.196.2013,mukherjee.nc.8.13918.2017,maczewsky.nc.8.13756.2017,ivanov.apl.4.126101.2019,arkhipova.sb.68.2017.2023,zhong.pr.2024}, 
valley Hall~\cite{noh.prl.120.063902.2018}, and higher-order topological insulators~\cite{hassan.np.13.697.2019,kirsch.np.17.995.2021}.  
Moreover, the photonic systems can also be nonlinear, affording the possibility to explore the interplay between geometry and nonlinearity~\cite{kartashov.prl.128.093901.2022}. Lattice solitons are extensively studied in periodic and aperiodic lattices with Kerr nonlinearity~\cite{ablowitz.pra.90.023813.2014,leykam.prl.117.143901.2016,maczewsky.science.370.701.2020,mukherjee.science.368.856.2020,ivanov.acs.7.735.2020,arkhipova.sb.68.2017.2023,ren.light.12.194.2023} and the nonlinear properties within this unique ``decorated boundary'' system remain largely uncharted.

In this Letter, we confirm the emergence of chiral bulk solitons within a photonic graphene structure subject to nonlinearity. By solving the nonlinear Schr\"odinger equation, the nonlinear chiral bulk states are found in the exact form of stationary nonlinear solutions bifurcating from the linear chiral bulk states.  Despite their formation, these nonlinear chiral bulk states are susceptible to dynamical instabilities. By employing the quasi-soliton approximation, we derive the analytical form of the envelope solution, thereby achieving the chiral bulk soliton. The chiral bulk soliton exhibits remarkable stability and the energy does not decay even after a long-time evolution. This discovery paves the way for a new paradigm in the manipulation of bulk-state transport, enabling control over these states without extensive bulk modifications. Our findings have significant implications for the design of photonic devices and offer a strategy for achieving the desired transport through the decorated method. Furthermore, these insights extend beyond the realm of photonics, enriching the broader field of topological materials and inspiring material design and functionality.

%\textcolor{red}{Last but not least, we would like to give an explanation on the topological protection (which is chiral protection here) phenomenon in Fig.~\ref{fig4} based on the tight-binding method. According to the Hamiltonian of the decorated honeycomb lattice in Fig.~\ref{fig1}(b), the Zak phase~\cite{delplace.prb.84.195452.2011} of the chiral bulk state can be calculated based on the Wilson-loop method~\cite{wang.njp.21.093029.2019,ren.pra.107.043504.2023}, which is $\pi$ when the chiral bulk state is in the band gap~\footnote{See Supplemental Materials for calculation details on the Zak phase}.}

\begin{figure}[htp!]
\centering
\includegraphics[width=1\columnwidth]{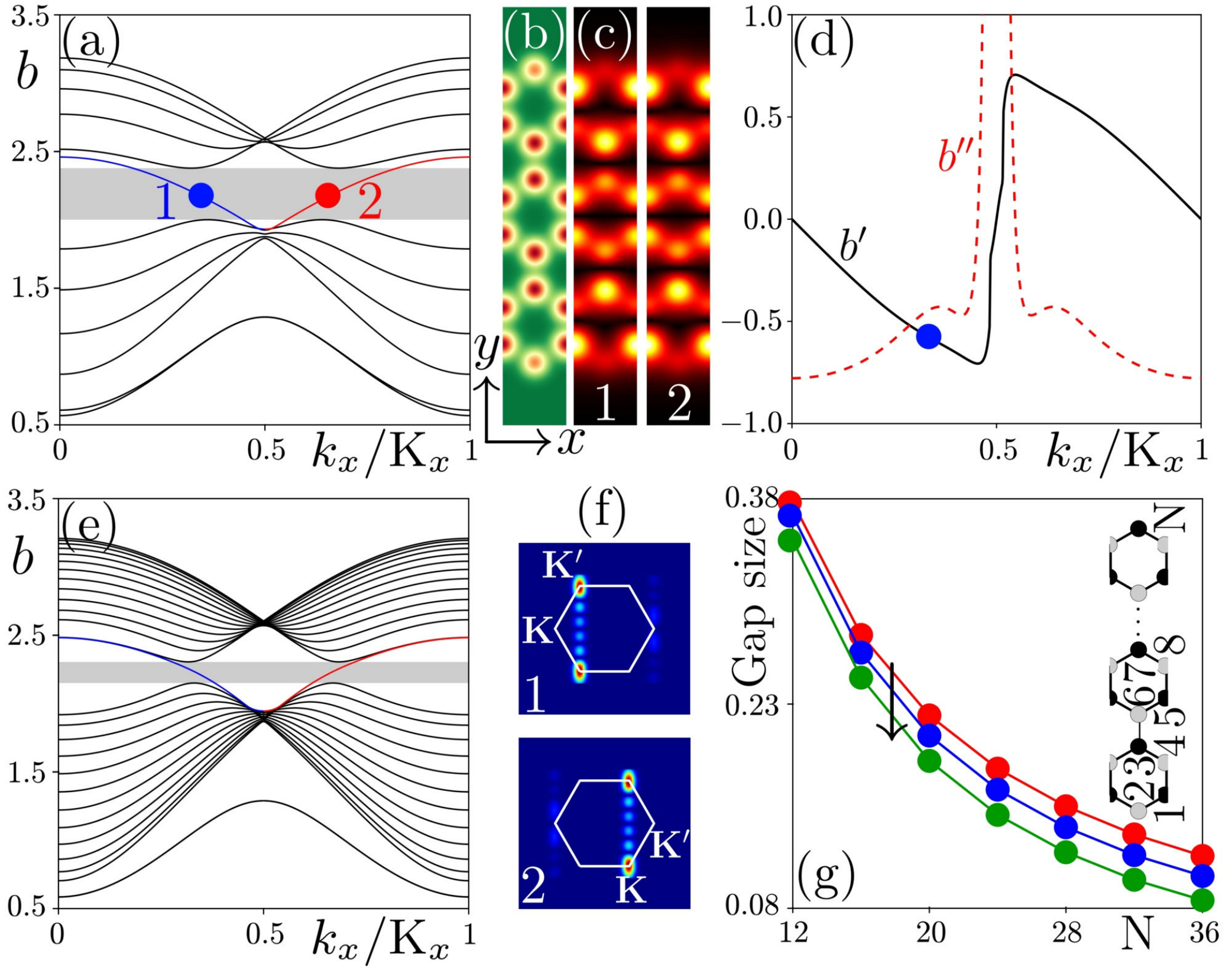}
\caption{(a) Band structure of the honeycomb lattice ribbon with 12 sites in the supercell. 
The red and blue curves represent different chiral bulk states. The shaded region represents the band gap.
(b) Honeycomb supercell. The upper and bottom sites are shallower than other sites. 
(c) Field modulus distribution of the chiral bulk state at ${k_x=0.33 {\rm K}_x}$ and ${k_x=0.67 {\rm K}_x}$ that are corresponding to the blue (numbered 1) and red (numbered 2) dots in (a), respectively. 
(d) First-order $b'$ (black solid curve) and second-order $b''$ (red dashed curves) derivatives of the chiral bulk state.
(e) Band structure of the honeycomb lattice ribbon with 28 sites in the supercell. 
(f) Fourier spectra of the chiral bulk states at dots 1 and 2 in (a). The white hexagon indicates the first Brillouin zone.
(g) Width of the band gap depends on the number $n$ of lattice sites in the supercell. 
The red, blue, and green curves correspond to the depth ${{\bm p}=(8,6)}$, $(8,4)$, and $(8,2)$ respectively.
Panels in (b,c) are shown in the window ${0\le x \le \sqrt{3}a}$ and ${-8\le y \le 8}$.}
\label{fig1}
\end{figure}

The propagation dynamics of a light beam in waveguide arrays with focusing cubic nonlinearity can be depicted by the Schr\"odinger-like paraxial wave equation
\begin{align}\label{eq1}
i \frac{\partial \psi}{\partial z} = -\frac{1}{2} \left( \frac{\partial^2}{\partial x^2} + \frac{\partial^2}{\partial y^2} \right) \psi -\mathcal{R}(x,y) \psi-|\psi|^{2} \psi,
\end{align}
where the transverse coordinate $x,y$ are normalized to the characteristic scale $r_0$
the longitudinal coordinate $z$ is normalized to the diffraction length $kr_0^2$,
${k=2n\pi/\lambda}$ is the wavenumber, $n$ is the background refractive index, and $\lambda$ is the wavelength.
In Eq.~(\ref{eq1}), the term $\mathcal{R}(x,y)$ represents the lattice waveguide array with the transverse landscape described by the Gaussian functions
$
{\mathcal{R} (x,y) =  p \sum_{m,s}  e^{- [(x-x_{m,s})^2+ (y-y_{m,s})^2]/\sigma^2} ,}
$
where ${p=k^2r_0^2\delta n/n}$ is the lattice depth with $\delta n$ being the refractive index change and
$\sigma$ is the width of the lattice site.
The typical values for the aforementioned waveguides are ${r_0=10\,\mu \rm m}$, ${\lambda=800\,\rm nm}$, ${n=1.45}$,
${\delta n \approx 1.1\times 10^{-4}}$ for ${p=1}$, ${\sigma=0.5}$ (corresponding to ${5\,\mu \rm m}$), and
the lattice constant ${a=1.4}$ (corresponding to ${14\,\mu \rm m}$).
In this Letter, we adopt the honeycomb lattice with zigzag-zigzag boundaries~\cite{plotnik.nm.13.57.2014} in $y$
and being periodic in $x$, as shown in Fig.~\ref{fig1}(b).
We set ${p=8}$ for most of the sites, while ${p=6}$ for the uppermost ($s=1$) and the nethermost ($s=N$) sites.
For convenience, we use the binary vector symbol ${{\bm p}=(8,6)}$ to represent the depth of the lattice:
the second number is the depth of the uppermost and nethermost sites of the supercell, 
while the first number is the depth of the other sites.
This operation opens the Dirac cones in the band structure of the honeycomb lattice to form valleys. 

By using the plane wave expansion method, the band structure of the lattice can be obtained 
if we assume the ansatz for Eq.~(\ref{eq1}) to be ${\psi=u(x,y) e^{ibz}}$ with $b$ being the propagation constant
and $u(x,y)$ the Bloch wavefunction that meets the condition ${u(x,y)=u(x+\ell X,y)}$,
by neglecting the nonlinear term.
Here, ${X=\sqrt{3}a}$ is the period in $x$ and $\ell$ is an integer.
The band structure, $b$ as a function of the Bloch momentum $k_x$, is shown in Fig.~\ref{fig1}(a),
in which the black curves are the bulk states, while the blue and red curves are the chiral bulk states
which are in the band gap indicated by a gray stripe.
Note that the width of the Brillouin zone is ${{\rm K}_x=2\pi/X}$.
We choose two chiral bulk states from the blue and red branches, respectively,
as indicated by the blue dot numbered 1 and the red dot numbered 2,
and show their field modulus profiles in Fig.~\ref{fig1}(c), which look the same but their phase distributions are distinct (not shown here).
The energy of the states occupies all the sites in the supercell, indicating that this in-gap chiral state is a bulk state.
We also show the first-order ${b'=db/dk_x}$ and second-order ${b''=d^2b/dk_x^2}$ derivatives of the chiral bulk states,
and show them in Fig.~\ref{fig1}(d), by the solid and dashed curves, respectively.
The signs of the blue and red chiral bulk states are opposite, 
which means that the bulk states of the blue branch and that of the red branch move in positive $x$ and negative $x$ during propagation, according to the relationship between the moving velocity $v=b'$ and $v=-b'$.
The second-order derivative $b''$ is nearly negative in the whole first Brillouin zone,
so one can expect that bright chiral bulk solitons are supported in this system~\cite{ivanov.acs.7.735.2020,ivanov.ol.45.1459.2020,ivanov.ol.45.2271.2020,tang.oe.29.39755.2021,zhong.ap.3.056001.2021,ren.nano.10.3559.2021,tang.chaos.161.112364.2022}.
By the way, if the uppermost and nethermost sites have deeper depth than other sites,
one may obtain dark chiral bulk solitons~\footnote{See Supplemental Materials for the detials on the dark chiral bulk solitons.}.
In addition to the field modulus distributions shown in Fig.~\ref{fig1}(c), the corresponding Fourier spectra are shown in Fig.~\ref{fig1}(f).
The energy scatters symmetrically to two equivalent $\bf K$ (${\bf K}'$) points, showing the valley-dependent behavior of these bulk modes. If we increase the number of unit cells to 28 sites,
the band structure is shown in Fig.~\ref{fig1}(e).
One finds that the chiral bulk states are still there, but the band gap becomes much narrower.
Figure~\ref{fig1}(g) collects the dependence of the width of the band gap on the site number $N$ in the supercell as well as
on the depth $\bm p$.
With an increasing number of sites, the band gap shrinks quickly and finally may disappear if the sites are more and more.
As to the depth, it affects the width of the band gap weakly even though the larger the depth is the wider the band gap is.
Therefore, one has to choose a relatively small unit cell with deeper depth to avoid the finite-size effect as far as possible.

\begin{figure}[htp!]
\centering
\includegraphics[width=1\columnwidth]{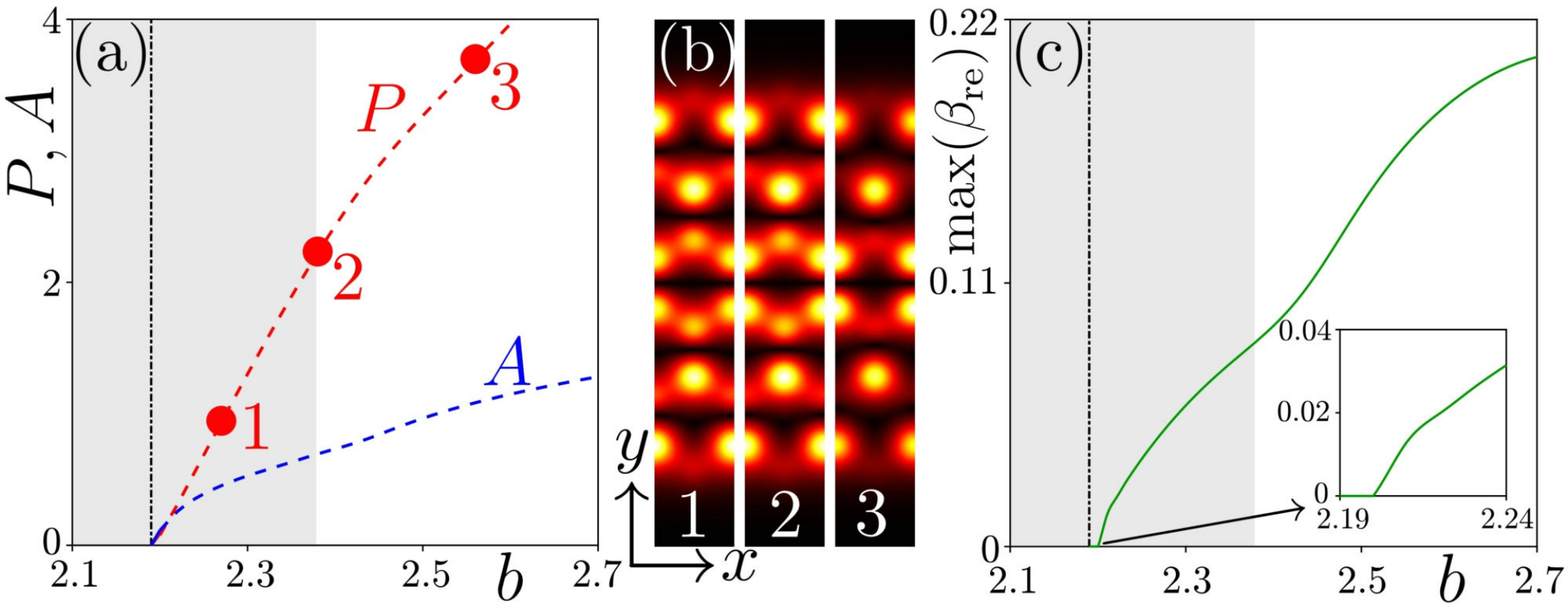}
\caption{(a) Nonlinear chiral bulk states family bifurcating from linear mode. The gray region is the band gap and the vertical dotted line shows the location of the linear chiral bulk state corresponding to the dot 1 in Fig.~\ref{fig1}(a).
$P$ is the power and $A$ is the peak amplitude.
(b) Maximum of the real part of the perturbation growth rate $\beta$ vs propagation constant $b$ of the nonlinear chiral bulk states in (a). The inset is a magnified plot close to the vertical dotted line.}
\label{fig2}
\end{figure}

Starting from the linear chiral bulk state, we can seek the corresponding nonlinear chiral bulk states
by solving Eq.~(\ref{eq1}) with the Newton method.
The power ${P=\iint |\psi|^2 dx dy}$ (red curve) and the peak amplitude ${A=\max\{|\psi|\}}$ (blue curve) of the nonlinear chiral bulk state are shown in Fig.~\ref{fig2}(a).
It is observed that $P$ of the non-linear chiral bulk state exhibits almost linear behavior with respect to its propagation constant $b$, while $A$ does not.
As the power of the nonlinear chiral bulk state decreases, the power curve and the peak amplitude curve gradually converge and finally intersect at ${b=2.19}$, which is the propagation constant of the linear chiral bulk state [blue dot numbered 1 in Fig.~\ref{fig1}(a)], as indicated by a vertical dotted line.
This illustrates that the nonlinear term in Eq.~(\ref{eq1}) can be neglected if the power of the state is sufficiently small. Therefore, the excitation of the nonlinear chiral bulk state is power thresholdless, concluding that the nonlinear chiral bulk state bifurcates from its linear counterpart.
Three nonlinear chiral bulk states are selected as indicated by three dots numbered ${1\sim3}$ on the power curve in Fig.~\ref{fig2}(a),
and their field modulus profiles are shown in Fig.~\ref{fig2}(b). The nonlinear chiral bulk states, whether located within the band gap (dot 1), within the bulk band (dot 3), or at the boundary between these regions (dot 2), exhibit remarkably similar mode profiles. This suggests that the nonlinear chiral bulk state does not hybridize with the other bulk states, despite their spatial overlap.

To check the stability of the nonlinear chiral bulk state, we use the stability analysis method, by introducing small perturbations $v(x,y)$ and $w(x,y)$ to the general solution of Eq.~(\ref{eq1}): $\psi=[u(x,y)+v(x,y)e^{\beta z}+w^*(x,y)e^{\beta^*z}]e^{i(bz+k_{x}x)},$ with $\beta$ being the perturbation growth rate and the asterisk as conjugate operation. By substituting this perturbed solution into Eq.~(\ref{eq1}), we derive a linear eigenvalue problem to determine $\beta$ for each nonlinear chiral bulk state. A state is unstable if the real part of $\beta$ is positive. Surprisingly, almost all nonlinear chiral bulk states are unstable except some very close to the linear chiral bulk states, shown as the relationship between $\max\{ \beta_{\rm re} \}$ (i.e., the maximum of the real part of $\beta$) and $b$ in Fig.~\ref{fig2}(c) and its inset where ${b=2.19}$ is the propagation constant of the linear chiral bulk state.
However, as shown in the following, one can still obtain a variety of stable chiral bulk solitons, which is expected, too.   

\begin{figure}[htbp]
\centering
\includegraphics[width=1\columnwidth]{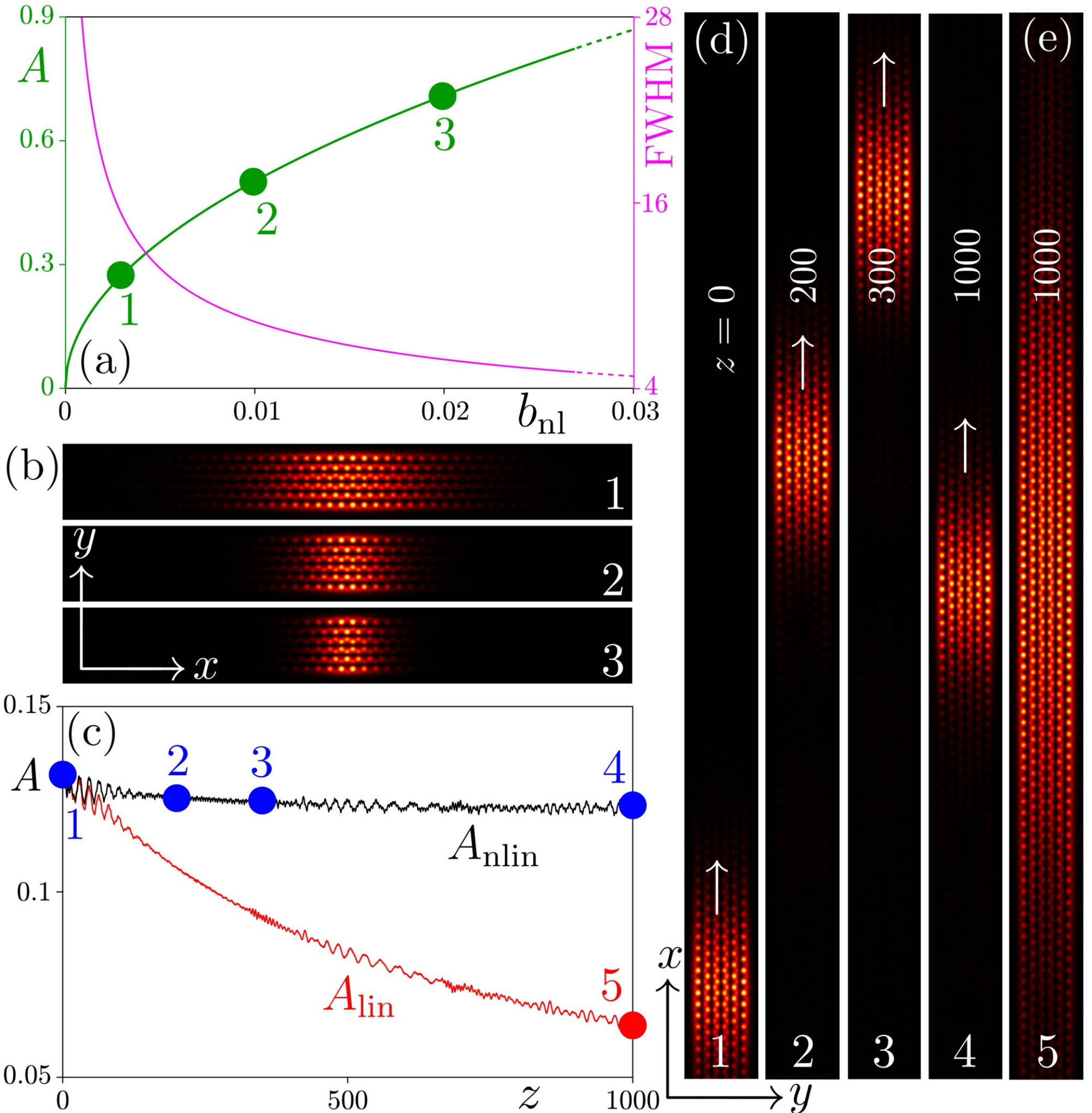}
\caption{(a) Dependence of the peak amplitude $A$ on the $b_{\rm nl}$ for the soliton family (green curve and refer to the left $y$ axis). Solid and dashed lines represent stable and unstable solitons, respectively. The FWHM as a function of $b_{\rm nl}$ is shown by the magenta curve (refer to the right $y$ axis).
(b) Amplitude modulus profiles of the chiral bulk soliton corresponding to the dots ${1\sim3}$ in (a).
(c) Peak amplitude of the soliton, corresponding to dot 1 in (a), during propagation. The black curve is for the nonlinear propagation, while the red curve records the linear propagation. Parameters are ${b_{\rm nl} = 0.003}$, ${k_x=0.33 {\rm K}_x}$, ${b'=-0.5723}$, ${b''=-0.4497}$ and ${\chi=0.0798}$.
(d) Amplitude modulus profiles of the chiral bulk soliton at typical distances corresponding to the dots ${1\sim4}$ in (c). The arrow indicates the moving direction of the soliton.
(e) Amplitude modulus profile of the soliton at ${z=1000}$ after linear propagation corresponding to the dot 5 in (c). Panels in (b) are shown in the window ${-60\le x \le 60}$ and ${-8\le y \le 8}$. Panels in (d,e) are shown in the window ${-115\le x \le 115}$ and ${-8\le y \le 8}$.}
\label{fig3}
\end{figure}

By employing the quasi-soliton solution, we can achieve the shape of chiral bulk solitons as 
$
\psi=\int_{-{\rm K}_x / 2}^{{\rm K}_x / 2} \mathcal{A}(\kappa, z)u(x, y, k+\kappa) e^{i bz+i(k+\kappa)x} d \kappa,    
$
here $\kappa$ is the momentum offset from the momentum $k$ with the amplitude $\mathcal{A}(\kappa, z)$ defined the envelope. Using Taylor series expansion at $\kappa$ for $u(x, y, k+\kappa)$, one can arrive at 
\begin{equation} \label{eq3}
\psi=e^{i bz+i k x} \sum_{n=0, \infty} \frac{(-i)^n}{n!}\frac{\partial^n u(x,y,k)} {\partial k^n}\frac{\partial^n \mathcal{A}(x, z)} {\partial x^n},
\end{equation}
where ${\mathcal{A}(x, z)=\int_{-\mathrm{K}_x / 2}^{\mathrm{K}_x / 2} \mathcal{A}(\kappa, z) e^{i \kappa x} d \kappa}$ is the envelope function of the corresponding nonlinear states. By choosing a gauge 
$
{\langle u(x,y,k), \partial_k u(x,y,k) \rangle=0,}
$
the system maintains the U(1) rotation of eigenvector phases and allows a valid assumption that ${n=0}$ for Eq.~(\ref{eq3}). Following the previous method~\cite{ivanov.acs.7.735.2020}, we obtain the slowly-varying envelope equation 
\begin{equation}
{i \frac{\partial \mathcal{A}} {\partial z}=\frac{b''}{2} \frac{\partial^2 \mathcal{A}}{\partial^2\xi}-\chi|\mathcal{A}|^2\mathcal{A},}
\end{equation}
%with the soliton solution expressed as ${\psi(x,y,z)=A(\xi,z)u(x,y)\exp(ibz)}$ for Eq.~(\ref{eq1}),
where ${\chi=\int_{-\infty}^{+\infty}dy\int_{0}^{X}|u|^{4}dx}$, and ${\xi= x+b' t}$. 
The analytical bright soliton solution for this envelope equation is
\begin{equation}
{\mathcal{A}=\sqrt{2 b_{\rm nl}/\chi} {\rm sech} (\xi \sqrt{-2 b_{\rm nl} / b''} ) \exp (-i b_{\rm nl} z ),}
\end{equation}
where $b_{\rm nl}$ is the propagation constant detuning which should not be large for stable solitons.
The full width at half maximum (FWHM) of the soliton intensity profile can also be analytically obtained as:
$\ln(\sqrt{2}+1)\sqrt{-2b''/b_{\rm nl}}$.
In Fig.~\ref{fig3}(a), the peak amplitude $A$ and the FWHM of the soliton envelope are displayed.
With increasing $b_{\rm nl}$, the peak amplitude increases while the FWHM decreases.
By superimposing the envelope to linear chiral bulk states, one obtains the chiral bulk solitons,
from which we choose three examples numbered ${1\sim3}$ (with ${b_{\rm nl}=0.003}$, $0.01$, and $0.02$, respectively) and show their field modulus distributions in Fig.~\ref{fig3}(b).
We introduce a small perturbation with amplitude upto $5\%$ of the soliton and do a long-distance propagation ${z\sim4000}$,
we find that the chiral bulk soliton becomes unstable if ${b_{\rm nl}>0.027}$, as indicated by the dashed lines in Fig.~\ref{fig3}(a).
The stable region for the chiral bulk solitons is quite large, since the width of the soliton should not be too narrow.
Therefore, one can always obtain stable chiral bulk solitons even though most nonlinear chiral bulk states are unstable.
Take the chiral bulk soliton numbered 1 as an example, we record its peak amplitude $A_{\rm nlin}$ during propagation according to Eq.~(\ref{eq1})
and show it in Fig.~\ref{fig3}(c); see the black curve.
One finds that the peak amplitude does not decay during propagation.
Field modulus profiles of the chiral bulk soliton at four selective distances, as indicated by the dots numbered ${1\sim4}$ on the black curve in Fig.~\ref{fig3}(a), are shown in Fig.~\ref{fig3}(d).
Indeed, the beam maintains its profile nearly unchanged during propagation in positive $x$ direction (remember the corresponding first-order derivative ${b'<0}$)~\footnote{Since the calculation window is limited, the beam will appear from the bottom end of the window in Fig.~\ref{fig3}(d) if it reaches the upper end according to the beam propagation method.}.
As a comparison, we lift the nonlinear term in Eq.~(\ref{eq1}) and check the linear propagation of the chiral bulk soliton.
The peak amplitude $A_{\rm lin}$ is indicated by the red curve in Fig.~\ref{fig3}(c) which decays with propagation distance and is distinct from the black curve.
The field modulus profile at ${z=1000}$, as shown by the dot numbered 5 in Fig.~\ref{fig3}(c), is exhibited in Fig.~\ref{fig3}(e)
which spreads greatly due to the diffraction without balancing with the focusing effect from the nonlinearity.
Until now, we can state that we create stable chiral bulk solitons.

%\begin{equation}\label{eq4}
%{\rm i}\beta \left[ \begin{array}{l}
%v\\
%w
%\end{array} \right] = \left[ {\begin{array}{*{20}{c}}
%{ - {{\cal L}_{{\rm{edge}}}} - 2|u{|^2}}&{ - |u{|^2}}\\
%{{{\left| u \right|}^2}}&{{{\cal L}_{{\rm{edge}}}} + 2{{\left| u \right|}^2}}
%\end{array}} \right]\left[ \begin{array}{l}
%v\\
%w
%\end{array} \right],
%\end{equation}
%where $\mathcal{L}_{\rm edge}=\frac{1}{2}(\frac{\partial}{\partial x^{2}}+\frac{\partial}{\partial y^{2}}+2{\rm i}k_{x}\frac{\partial}{\partial x}+k_{x}^{2})+(\mathcal{R}-b)$.

%These chiral bulk states do not require (artificial) magnetic fields, which greatly facilitates their practical applications.

\begin{figure}[htp!]
\centering
\includegraphics[width=1\columnwidth]{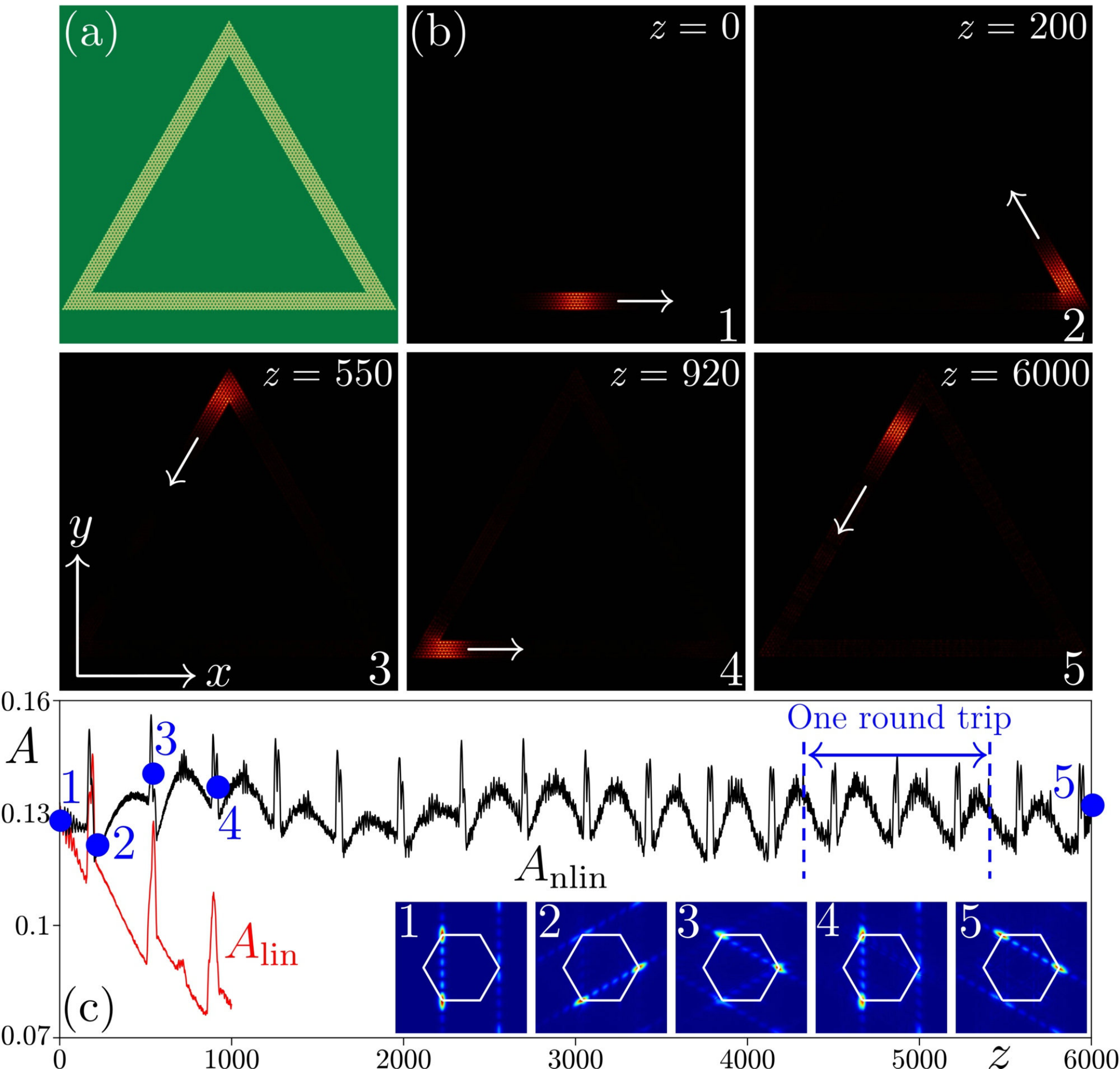}
\caption{(a) Triangular configuration of the lattice that has three sharp corners.
(b) Amplitude modulus profiles of the chiral bulk soliton at typical distances.
(c) Peak amplitude of the soliton during propagation (black curve). The red curve is its peak amplitude during linear propagation.
Dots are corresponding to the panels in (b). The panels below are the total spatial spectra at typical distances corresponding to the dots ${1\sim5}$.
Parameters are the same as those adopted in Fig.~\ref{fig3}(c). Panels in (a,b) are shown in the window ${-120\le x,y \le 120}$.}
\label{fig4}
\end{figure}

Therefore, it is interesting to inspect whether it can circumvent sharp corners as the valley Hall edge states.
To this end, the honeycomb lattice is elaborated to have three sharp corners, as shown in Fig.~\ref{fig4}(a).
The chiral bulk soliton corresponding to dot 1 in Fig.~\ref{fig3}(a) is launched into the bottom edge of the honeycomb lattice,
and its field modulus profiles during its counter-clockwise propagation at typical distances are shown in Fig.~\ref{fig4}(b).
One finds that it can circumvent sharp corners without backscattering happening and profile changing; see the panels with ${z=200}$, ${z=550}$, and ${z=920}$ in Fig.~\ref{fig4}(b).
This property holds even though the propagation distance reaches ${z=6000}$.
The peak amplitude of the soliton ${A_{\rm nlin}}$ during nonlinear propagation, as illustrated by the black curve in Fig.~\ref{fig4}(c),
demonstrates this capability of the chiral bulk soliton in a clearer way: The peak amplitude oscillates but does not decay after rotating more than 5 rounds.
The five soliton beams exhibited in Fig.~\ref{fig4}(b) are marked on the peak amplitude curve with dots numbered from 1 to 5. corresponding to the five solitons beams in Fig.~\ref{fig4}(b),
the spectra are displayed in Fig.~\ref{fig4}(c) as insets.
One finds that the soliton beam is always at the $\bf K$ points, 
which demonstrates that the inter-valley scattering indeed does not happen.
As a comparison, we also investigate the linear propagation of the same chiral bulk state in this triangular configuration.
The peak amplitude ${A_{\rm lin}}$ is shown by the red curve in Fig.~\ref{fig4}(c),
which decreases dramatically.

To address this challenge, we introduce a chiral bulk soliton in photonic graphene with decorated boundaries. By incorporating nonlinearity into the system, we demonstrate chiral bulk solitons in a photonic graphene lattice. This discovery suggests an efficient method for manipulating the transport of bulk modes, without the extensive and often complex bulk modifications. Our findings not only contribute to the fundamental understanding of nonlinear photonic systems but also offer a promising avenue for the design and optimization of photonic devices. The extensive functionalization of the bulk area will open a revolutionary era in optical communication~\cite{agrell_jo.18.063002.2016}, quantum computing~\cite{brien.science.318.1567.2007,pelucchi.nrp.4.194.2022}, and topological lasing~\cite{shao.nn.15.67.2020,wong.prr.3.033042.2021}. 

The authors acknowledge Ruoyang Zhang for discussions. This work was supported by the Natural Science Basic Research Program of Shaanxi Province (2024JC-JCQN-06), the National Natural Science Foundation of China (12074308), and the Fundamental Research Funds for the Central Universities (xzy022023059). 

% Bibliography
%merlin.mbs apsrev4-1.bst 2010-07-25 4.21a (PWD, AO, DPC) hacked
%Control: key (0)
%Control: author (72) initials jnrlst
%Control: editor formatted (1) identically to author
%Control: production of article title (-1) disabled
%Control: page (0) single
%Control: year (1) truncated
%Control: production of eprint (0) enabled
%

\end{document}